\documentclass[12pt]{article}
\usepackage{amsfonts}

\def\boldsymbol#1{\setbox0=\hbox{$#1$}%
 \kern-.025em\copy0\kern-\wd0
 \kern.05em\copy0\kern-\wd0
 \kern-.025em\raise.0433em\box0 }

\title{Minkowski electromagnetic analog for Kerr-Newman electrovac solution}
\author{S M Blinder\\University of Michigan\\Ann Arbor, MI 48109-1055, USA
\\sblinder@umich.edu}

\begin{document}

\maketitle 

\begin{abstract}

\noindent An electromagnetic analog of the Kerr-Newman solution in general relativity is
derived, based on Minkowski's formulation for electromagnetic fields in moving media. The
equivalent system is a distribution of charges and currents largely localized within a
spinning disk of radius $a$.  This occurs in a rotating medium of inhomogeneous index
of refraction and ``frame dragging" at exactly half
the angular velocity of the electrical charge.

\end{abstract}

% PACS 04.20.-q,03.50.De}

\section{Introduction}

There has been much current interest in simulating experiments on black holes
and related exotic astrophysical phenomena which are currently inaccessible
(or inadvisable) by using analog models in condensed matter\cite{1}. 
For example, optical analogs of light-trapping black holes can be created, in
concept, by a combination of extremely high index of refraction and dielectric
fluid flow. One realization has been the extreme subluminal 
propagation of light through Bose-Einstein condensates\cite{2}.

In this paper we will derive an exact analog in Lorentzian spacetime for the
electromagnetic properties of a Kerr-Newman black hole, based on
Minkowski's 1908 formulation of relativistic electrodynamics in moving
dielectric media\cite{m}.

\section{Kerr-Newman Geometry} 

Kerr\cite{ke} first solved Einstein's equations
for a spinning black hole of mass $M$ and angular momentum per unit
mass $a= S/M$.  Newman\cite{ne} generalized this result for a black hole carrying electric
charge
$Q$.  The metric for Kerr-Newman geometry, in the spherical coordinates $\{t, r, \theta,
\phi\}$ introduced by Boyer and Lindquist\cite{bl}, can be written
\begin{equation}  g_{\lambda\mu}=\left(\begin{array}{llll}
(a^2\sin^2\theta-\Delta)/\rho^2& 0 & 0
&-a(a^2+r^2-\Delta)\sin^2\theta/\rho^2\cr
0 & \rho^2/\Delta  & 0 & 0 \cr 
0 & 0 & \rho^2 & 0  \cr
-a(a^2+r^2-\Delta)\sin^2\theta/\rho^2 & 0 & 0 &
\left[(a^2+r^2)^2-a^2\Delta\sin^2\theta\right]\sin^2\theta/\rho^2
\end{array}\right)
\end{equation}
where
\begin{equation} \rho^2\equiv r^2+a^2\cos^2\theta
\qquad\qquad
 \Delta\equiv a^2+ r^2 -2Mr  +Q^2\end{equation}
The interval can be compactly represented as
\begin{equation}\qquad
 ds^2 =
-{\Delta\over\rho^2}\,\left[dt-a\,\sin^2\theta\,d\phi\right]^2+
{\sin^2\theta\over\rho^2}\,\left[(r^2+a^2)d\phi-a\,dt\right]^2
+{\rho^2\over\Delta}\,dr^2 +\rho^2\,d\theta^2
\end{equation}
The Jacobian for the metric (1) is given by
\begin{equation}
\sqrt{-g}=\rho^2\sin\theta=(r^2+a^2\cos^2\theta)\sin\theta\end{equation}

Solution of the Einstein-Maxwell equations for Kerr-Newman geometry gives an electrovac
stress-energy tensor with the following nonvanishing elements:
$T^\lambda_\mu$:
\begin{eqnarray}T^0_0= - T^3_3=-{
{Q^2(r^2+a^2+a^2\sin^2\theta)}\over
     {8\pi\, ( r^2 + a^2\,\cos^2\theta )^3 }}\qquad
 T^1_1=-  T^2_2=-{
{Q^2}\over
     {8\pi\, ( r^2 + a^2\,\cos^2\theta )^2 }}\cr\cr
 T^3_0=
-{{a\, Q^2}\over{4\pi \,
    {( r^2 +  a^2\,\cos^2\theta) }^3} }
\qquad
 T^0_3=
{{a\, Q^2\,
    \left( a^2 + r^2 \right) \,
    \sin^2\theta}\over{4\pi \,
    {( r^2 + 
        a^2\,\cos^2\theta
      ) }^3} }\qquad\qquad
\end{eqnarray}
Two of these tensor elements are directly related to the energy density and angular momentum
density, namely,
\begin{equation} {\cal W}=T^0_0\,\sqrt{-g},\qquad {\cal S}=\frac1c\, T^0_3\,\sqrt{-g}
\end{equation}
The asymptotic form  of the metric element $g_{03}=-2Ma\sin^2\theta
\,r^{-1}+{\cal O}(r^{-2})$ is also consistent with an angular momentum $Ma$.   

When $a=0$, 
the above result  reduces to the Reissner-Nordstr\o m  solution\cite{rn}.
In flat Lorentzian spacetime, obtained by setting $a=M=Q=0$, the metric (1) reduces to
\begin{equation}  
\eta_{\lambda\mu}=\left(\begin{array}{cccc}
-1 &  & &\bigcirc \cr & 1  & &\cr
& & r^2 &  \cr   \bigcirc &  &  & r^2\sin^2\theta
\end{array}\right)
\end{equation}

\section{Minkowski electrodynamics}

The fundamental equations of electrodynamics in curved spacetime are the following:
\begin{equation}\frac\partial{\partial x^\beta}\left( \sqrt{-g}\,F^{\alpha\beta}\right)
=\frac{4\pi}c\,
\sqrt{-g}\,j^\alpha
\end{equation}
\begin{equation}\epsilon^{\alpha\beta\gamma\delta}\frac\partial{\partial x^\gamma}
F_{\alpha\beta}=0
\end{equation}
\begin{equation}
T^\lambda_\mu=\frac1{4\pi}\,\left(F^{\lambda\alpha}F_{\mu\alpha}
-\frac14\,\delta^\lambda_\mu\,F^{\alpha\beta}F_{\alpha\beta}\right)
\end{equation}
Eqs (8) and (9) are generalizations of Maxwell's equations, while (10) gives the Maxwell
stress-energy tensor.

In Minkowski's formulation of relativistic electrodynamics for moving
media\cite{m}, Maxwell's equations take the form
\begin{equation}\frac\partial{\partial x^\beta} {\mathfrak G}^{\alpha\beta}
=\frac{4\pi}c\,
{\mathfrak j}^\alpha
\end{equation}
\begin{equation}\epsilon^{\alpha\beta\gamma\delta}\frac\partial{\partial x^\gamma}
{\mathfrak F}_{\alpha\beta}=0
\end{equation}
while the stress-energy tensor in a medium is generalized to
\begin{equation}
{\mathfrak  T}^\lambda_\mu=\frac1{4\pi}\,\left({\mathfrak  G}^{\lambda\alpha}{\mathfrak
F}_{\mu\alpha} -\frac14\,\delta^\lambda_\mu\,{\mathfrak  G}^{\alpha\beta}{\mathfrak
F}_{\alpha\beta}\right)
\end{equation}
Eqs (11)-(13) pertain to the flat Lorentzian spacetime of special relativity. The
corresponding relations in curved spacetime can evidently be obtained by the formal
replacments
\begin{equation}\quad{\mathfrak  j}^\lambda\to\sqrt{-g}\,j^\lambda ,\quad
{\mathfrak G}^{\lambda\mu}\to\sqrt{-g}\,F^{\lambda\mu}  ,\quad {\mathfrak
F}_{\lambda\mu}\to F_{\lambda\mu},\qquad  {\mathfrak T}^\lambda
_\mu\to\sqrt{-g}\,T^\lambda _\mu
\end{equation}

The covariant field tensor in spherical coordinates is given by
\begin{equation}  
{\mathfrak F}_{\lambda\mu}=\left(\begin{array}{cccc}
0 & E_r  & r\, E_\theta &  r\sin\theta\,E_\phi \cr 
-E_r & 0  &- r\, B_\phi & r\sin\theta\,B_\theta \cr
-r \,E_\theta & r\,B_\phi & 0 & -r^2\sin\theta\, B_r  \cr   
-r\sin\theta\,E_\phi &- r\sin\theta\,B_\theta  & r^2\sin\theta\, B_r & 0
\end{array}\right)
\end{equation}
while the contravariant auxilliary field tensor is
\begin{equation} 
{\mathfrak G}^{\lambda\mu}=\left(\begin{array}{cccc}
0 & -D_r  &  -D_\theta/r &  -D_\phi/r\sin\theta \cr 
D_r & 0  &- H_\phi/r & H_\theta/r\sin\theta \cr
 D_\theta/r & H_\phi/r & 0 & - H_r/r^2\sin\theta  \cr   
D_\phi/r\sin\theta & -H_\theta/r\sin\theta  &  H_r/r^2\sin\theta & 0
\end{array}\right)
\end{equation}

Minkowski's original definitions of the constitutive relations was based on a homogeneous,
isotropic medium moving at constant velocity.  A number of variants and extensions have
been proposed by many authors\cite{ce}.  In this paper we will assume that Minkowski's
formulation provides a valid \textsl{phenomenological} representation, even for nonuniform
motion of inhomogeneous media.
 
Assuming the equivalence of the Kerr-Newman tensor (5) to the Minkowski
tensor (13), we obtain the simultaneous equations
\begin{eqnarray}
8\pi\sqrt{-g}\,T^0_0=-D_rE_r-H_rB_r-D_\theta E_\theta-H_\theta B_\theta\cr\cr
8\pi\sqrt{-g}\,T^1_1=-D_rE_r-H_rB_r+D_\theta E_\theta+H_\theta B_\theta\cr\cr
4\pi\sqrt{-g}\,T^3_0=(E_\theta H_r- E_rH_\theta)/r\sin\theta\cr\cr
4\pi\sqrt{-g}\,T^0_3=(D_r B_\theta- D_\theta B_r)\, r\sin\theta
\end{eqnarray}
where cylindrical symmetry implies that the $\phi$-components of the electric and
magnetic fields vanish. The solution with the simplest functional forms is found to be
\begin{eqnarray}\qquad
D_r= \frac{Q\,
    \left( r^2 - 
      a^2\,{\cos ^2\theta }
      \right) }{
    {\left( r^2 + 
        a^2\,
         {\cos^2\theta }
        \right)^2 }}\qquad\qquad
E_r=  
\frac{Q\,\left( a^2 + r^2
      \right) \,
    \left( r^2 - 
      a^2\,{\cos ^2\theta }
      \right) }{r^2\,
    {\left( r^2 + 
        a^2\,
         {\cos^2\theta }
        \right)^2 }}
\cr\cr
D_\theta=E_\theta= -\frac{2 \,a^2\, Q\,
   \sin\theta \cos\theta}{
    {\left( r^2 + 
        a^2\,
         {\cos^2\theta }
        \right)^2 }}\qquad\qquad D_\phi=E_\phi=0\qquad\qquad
\cr\cr
B_r= \frac{2\,a\,Q\,r\,
\cos\theta }{
    {\left( r^2 + 
        a^2\,
         {\cos^2\theta }
        \right)^2 }}\qquad\qquad
H_r= \frac{2\,a\,Q\,
  (a^2 + r^2)\cos\theta }{r\,
    {\left( r^2 + 
        a^2\,
         {\cos^2\theta }
        \right)^2 }}
\qquad\qquad\cr\cr
B_\theta=H_\theta= \frac{ a\, Q\,   \left( r^2 - a^2\,\cos ^2\theta \right) 
   \sin\theta}{r\,
    {\left( r^2 + 
        a^2\,
         {\cos^2\theta }
        \right)^2 }}\qquad\qquad B_\phi=H_\phi=0\qquad
\end{eqnarray}
The $\bf E$ and $\bf B$ fields are consistent with the scalar and vector potentials
\begin{equation}  
\Phi=\frac{Q\,r}{r^2 + a^2\,\cos^2\theta}\qquad\qquad {\bf A}=\frac{Q\,a\sin\theta}{r^2
+ a^2\,\cos^2\theta}\,\boldsymbol{\hat\phi}
\end{equation}  

As $r\to \infty$, the electric field approaches that of a point charge, with higher-order
contributions consistent with an oblate spheroidal distribution of approximate  
radius $a$. The magnetic field shows the asymptotic behavior of a point dipole
of magnitude $\mu=Qa=QS/M$.  As noted by Carter\cite{ca}, this is corresponds to a
$g$-factor of 2, just as in Dirac's relativistic quantum theory.  Perhaps
this $g$-factor for spin is a \textsl{classical} relativistic effect, not dependent on quantum
mechanics\cite{b}.
 
The \textsl{total} charge density, including both free and induced charges, is given by
\begin{equation}
\rho(r,\theta,\phi)=\frac{1}{4\pi}\,\boldsymbol{\nabla\cdot}{\bf E}=
\frac{Q a^2 (r^2 -3 a^2\,\cos^2\theta)}{2\,\pi\, r(r^2 + a^2\,\cos^2\theta)^3}
\end{equation}
with a radial distribution function    
\begin{equation}
D(r)=\int_0^{2\pi}\,\int_0^\pi\,\rho(r,\theta,\phi)\,r^2\sin\theta\,d\theta\,d\phi=
\frac{2\, Q a^2\, r}{(r^2 + a^2)^2}
\end{equation}
Eqs (20) and (21) show a predominant localization of charge in the vicinity of a ring of
radius $a$ in the medial plane $\theta=\pi/2$. The analogous total current density is 
\begin{equation}
{\bf j}=\frac{c}{4\pi}\,\nabla\times{\bf B}=
\frac{Q a^3 c\,(r^2 -3 a^2\,\cos^2\theta)\sin\theta}{2\,\pi\, r^2(r^2 +
a^2\,\cos^2\theta)^3}
\,\boldsymbol{\hat\phi}
\end{equation}
Note that the charge goes negative and current reverses where $r<\sqrt3\,a\cos\theta$.
It can be rationalized that this is what causes the enhancement of the $g$-factor.    
For larger values of $r$, the charge and current densities are related by
\begin{equation}
\qquad{\bf j}=\rho\,{\bf v}\qquad{\rm with} \qquad {\bf
v}=\frac{a\,c\sin\theta}{r}\,\boldsymbol{\hat\phi}
\end{equation}
This represents a convective current caused by rotation of the charge
distribution with angular velocity
\begin{equation}\qquad\qquad  \Omega={a\,c}/{r^2}      \end{equation}
around the polar axis, such that ${\bf v}=\boldsymbol{\Omega}\times{\bf r}$ 
This description of the source of the Kerr-Newman metric is in
essential agreement with the analysis of Pekeris and Frankowski\cite{pf}.

\section{Constitutive relations and frame dragging}

The electric permittivity and magnetic permeability consistent with the
fields (18) could be most simply accounted for by an anisotropic index of refraction,
differing in the radial and tangential directions, such that
\begin{equation}
\epsilon_{\rm r}=\mu_{\rm r}=n_{\rm r}=\frac{r^2 +
a^2}{r^2}\qquad\qquad\epsilon_{\rm t}=\mu_{\rm t}=n_{\rm t}=1
\end{equation}
More reasonable conceptually, however, is a model based on
an isotropic but \textsl{rotating} medium. Minkowski defined $\epsilon$ and $\mu$ in a
moving medium such that
\begin{eqnarray}
{\bf D}+\frac1c\,({\bf v\times H})=\epsilon\left[{\bf E}+\frac1c\,({\bf
v\times B})\right]
\cr\cr\qquad
{\bf B}-\frac1c\,({\bf v\times E})=\mu\left[{\bf H}-\frac1c\,({\bf
v\times D})\right]
\end{eqnarray}
Solving for ${\bf D}$ and ${\bf B}$, we find
\begin{eqnarray}
{\bf D}=\frac{\epsilon(1-\beta^2){\bf E}+(\epsilon\mu-1)\boldsymbol{\beta}\times
{\bf H}} {1-\epsilon\mu\beta^2}
\cr\cr\qquad
{\bf B}=\frac{\mu(1-\beta^2){\bf H}-(\epsilon\mu-1)\boldsymbol{\beta}\times{\bf
E}} {1-\epsilon\mu\,\beta^2}
\end{eqnarray}
where
\begin{equation}  {\boldsymbol{\beta}}=\frac{\bf v}c=\frac{
\boldsymbol{\omega}\times{\bf r}}{c}  =\frac{\omega\, r
\sin\theta}{c}\,\boldsymbol{\hat\phi} \end{equation} 
with angular velocity
$\omega$ again around the polar axis.   Eqs  (18) and (27) are
consistent with

\begin{equation}\qquad\epsilon=\mu=n=\frac{a^4 + 2\,r^2(r^2 +  a^2\,
    \cos^2\theta)  + a^2\,{\sqrt{a^4 + 4\,r^2\,( r^2 + a^2\,
          \cos^2\theta 
           ) }}}{2\,r^2\,
   ( r^2 + 
      a^2\,\cos^2\theta)  }\end{equation}
and
\begin{equation}\beta=\frac{ a^2 + 2\,r^2 - 
      {\sqrt{a^4 + 
          4\,r^2\,
        ( r^2 + 
           a^2\,
          \cos^2\theta
          ) }}  }{2\,a\,r\,\sin\theta}\end{equation}

For $r\gg a$,
we find the limiting dependence
\begin{equation} \epsilon=1 + \frac{a^2}{r^2} + 
  \frac{\sin^2\theta \,a^4}{2\,r^4}+{\cal O}(r^{-6})
\end{equation}
and
\begin{equation}\beta=
\frac{\sin \theta\,a }{2\,r} - 
  \frac{\left( 1 + 
       \cos ^2\theta  \right)
       \,\sin\theta \, a^3}{8\,r^3} +
  {\cal O}(r^{-5})
\end{equation}
or
\begin{equation}
 \omega=\frac{a\, c}{2 \,r^2}\left[1-\frac{(1+\cos^2\theta)a^2}{4r^2}+{\cal
O}(r^{-4})\right]
\end{equation}
At the opposite limit, as $r\to 0$, 
\begin{equation}
\epsilon\approx\frac{a^2}{r^2\cos^2\theta}\quad {\rm and}
\quad\displaystyle \omega\approx\frac{c}{a}-\frac{(1+\cos^2\theta)c\, r^2}{a^3}
\end{equation}
The effective
Minkowski medium thus behaves like a vortex around the charged source with maximum
angular velocity
$c/a$ at the center, falling off asymptotically as $ac/2r^2$. The vorticity of the
rotating medium, $\boldsymbol{\Omega}=\nabla\times{\bf v}$, approaches a
limiting value twice that of $\omega$, namely  $\Omega\approx ac/r^2$. Remarkably, this
limiting vorticity is equal to the angular velocity of the circulating charge distribution given
by (24).

One might visualize an electromagnetic \textsl{frame dragging}, analogous to the
Lense-Thirring effect for gravitation, at \textsl{half} the angular velocity of the rotating charge
distribution. This might somehow be the basis for rationalization of the multiples of 2
occurring in the Dirac
$g$-factor, as well as the Thomas precession.


\begin{thebibliography}{99}

\bibitem{1} See, for example, Workshop on Analog Models of General Relativity
(Rio de Janeiro, October 2000):
{\tt http://www.physics.wustl.edu./ ${\scriptstyle\sim}$visser/Analog/bibliography.html}

\bibitem{2}  Hau L V \textsl{et al} \textsl{Nature} {\bf 397} 594 (1999); \textsl{Nature} {\bf 409}
490 (2001)

\bibitem{m} Minkowski H \textsl{Math. Ann.} {\bf 68} 472 (1910); summarized
in Pauli W \textsl{Theory of Relativity} (Pergamon, New York, 1958), pp 99 ff.


\bibitem{ke} Kerr R P  \textsl{Phys. Rev. Lett.}  {\bf 11} 237 (1963)

\bibitem{ne}
Newman  E T \textsl{et al}  \textsl{J. Math. Phys.}  {\bf 6}
918 (1965)

\bibitem{bl} Boyer  R H and Lindquist R W  \textsl{J. Math. Phys.} {\bf 8}, 2651
(1967)

\bibitem{rn} Reissner H \textsl{Ann. Phys. (Germany)} {\bf 50} 106 (1916);
Nordstr\o m G  \textsl{Proc. Kon. Ned. Akad. Wet.} {\bf 20} 1238
(1918)

\bibitem{ce} For a recent review, see Censor D, \textsl{Progress In Electromagnetics Research}
{\bf 29} 107 (2000).


\bibitem{ca} Carter B \textsl{Phys. Rev.}  {\bf 174} 1559 (1968)

\bibitem{b} Blinder S M  \textsl{Repts.
Math. Phys.}  {\bf 47} 279 (2001); Online version:  

{\tt http://arXiv.org/find/math-ph/1/au:+Blinder/0/1/0/past/0/1}

\bibitem{pf} Pekeris C L and Frankowski K \textsl{Phys. Rev. A}  {\bf 36} 5118 (1987)

\end{thebibliography}
\end{document}